\documentclass[3p,twocolumn]{elsarticle}

\usepackage{lineno,hyperref,pifont,natbib,geometry,fleqn,graphicx,makeidx,fancyhdr,multirow,verbatim,amsmath,amsthm,amssymb,float,array,epstopdf}
\modulolinenumbers[5]

\journal{Journal of \LaTeX\ Templates}

\begin{document}

\begin{frontmatter}

% \title{Elsevier \LaTeX\ template\tnoteref{mytitlenote}}
\title{Silicon Sensors for Trackers at High-Luminosity Environment}

\author{Timo Peltola\corref{cor1}}
\ead{timo.peltola@helsinki.fi}
\address{Helsinki Institute of Physics, P.O.Box 64 (Gustaf H\"{a}llstr\"{o}min katu 2) FI-00014 University of Helsinki, Finland}
\author{\small On behalf of the RD50 Collaboration\fnref{myfootnote}}
% \fntext[myfootnote]{Speaker.}
\fntext[myfootnote]{A complete author list can be found at: http://rd50.web.cern.ch/rd50.}
\cortext[cor1]{Speaker}

\begin{abstract}
% This template helps you to create a properly formatted \LaTeX\ manuscript.
\label{Abstract}
The planned upgrade of the LHC accelerator at CERN, namely the high luminosity (HL) phase of
the LHC (HL-LHC foreseen for 2023), will result in a more intense radiation environment than the present 
tracking system was designed for. The required upgrade of the all-silicon central trackers at the ALICE, ATLAS, CMS and LHCb experiments
will include higher granularity and radiation hard sensors. The radiation hardness of the new sensors must be roughly an order of magnitude
higher than in the current LHC detectors. To address this, a massive R$\&$D
program is underway within the CERN RD50 collaboration "Development of Radiation Hard Semiconductor Devices for Very High Luminosity Colliders" 
to develop silicon sensors with sufficient radiation tolerance.
Research topics include the improvement of the intrinsic radiation tolerance of the sensor material and novel detector designs
with benefits like reduced trapping probability (thinned and 3D sensors), maximized sensitive area (active edge sensors) and
enhanced charge carrier generation (sensors with intrinsic gain). A review of the recent results from both measurements and TCAD simulations of several detector technologies and silicon
materials at radiation levels expected for HL-LHC will be presented. 

\end{abstract}


\end{frontmatter}

% \linenumbers

\section{Introduction}
\label{Introduction}
Position sensitive silicon detectors are largely employed in the tracking systems of High Energy
Physics experiments due to their outstanding performance. They are currently installed in the
vertex and tracking part of the ALICE, ATLAS, CMS and LHCb experiments at LHC, the world's
largest particle physics accelerator at CERN.

An upgrade of LHC accelerator is already planned for 2023, namely the high luminosity phase of
the LHC (HL-LHC). Approximately a 10-fold increase of the luminosity will enable the use of maximal physics potential of the machine.
After 10 years of operation, the integrated luminosity of 3000 fb$^{-1}$ \cite{Rossi2012,Ruggiero2002,Gianotti2002}
will expose the tracking system at HL-LHC to a radiation environment that is beyond the capability of the present system design. At ATLAS and CMS fluences of more than $1\times10^{16}$ $\textrm{n}_{\textrm{\small eq}}$cm$^{-2}$ (1 MeV neutron equivalent) are expected for the pixel detectors at the innermost layers and fluences above $10^{15}$ $\textrm{n}_{\textrm{\small eq}}$cm$^{-2}$ for the strip sensors $\sim$20 cm from vertex \cite{Dawson2006}. Besides these extremely
high radiation levels the increase in the track density will be among the most demanding challenges. % and the proposed reduction of the bunch crossing time from 25 ns to about 10 ns will pose the severest challenges.
% Faster detectors with finer granularity are therefore required. 
Detectors with finer granularity are therefore required. This means e.g. that the pixel detectors will have to cover the tracking volume to higher radii moving the microstrip detector system further outward. 
The increased granularity will call for more cost effective technologies than presently existing since otherwise a
detector upgrade could not be afforded.

This requires a dedicated R$\&$D program to find solutions that improve the present detector technologies, or develop novel ones,
for the innermost tracking layers and for replacements for most of the outer detector components with detectors that can withstand higher radiation levels and higher occupancies.

The RD50 collaboration "Development of Radiation Hard Semiconductor Devices for Very High Luminosity
Colliders" \cite{theRD50} was formed in 2002 with the aim to develop
semiconductor sensors matching the above mentioned
HL-LHC requirements. About 280 members from 50 institutes and of several experiments at
the LHC are pursuing this aim in specific research fields, namely defect/material characterization, detector characterization, new structures and full detector 
systems. The collaboration arranges common sensor production runs, biannual
workshops and access to irradiation facilities. These efforts are now closely linked
to the increasing upgrade activities of ATLAS and CMS
as the RD50 results are already setting some baseline detector concepts for the upgrade.

In the following selected recent results of the RD50 scientific program
will be described. More detailed information can be found in Refs. \cite{theRD50,RD50sr2010} and in literature cited there.

% \section{Front matter}
\section{Defect and Material Characterization}
\label{Defect Characterization}
The main objective of the research field is the identification of the key radiation effects that adversely affect the sensor properties. Various techniques used for the defect analysis include thermally stimulated 
current technique (TSC), deep level transient spectroscopy (DLTS), transient current technique (TCT \cite{Eremin1996TCT}) or capacitance and leakage
current measurements versus depletion voltage, (C-V) and (I-V), respectively. The defects are characterized by their capture cross section, ionization energy, concentration and type.

The three radiation effects on the sensor bulk material are the increase of the leakage current, the trapping of charge carriers and the change of the effective doping concentration of the sensor and thus, the depletion voltage. 

\begin{figure}[tbp] 
\centering 
\includegraphics[width=.45\textwidth]{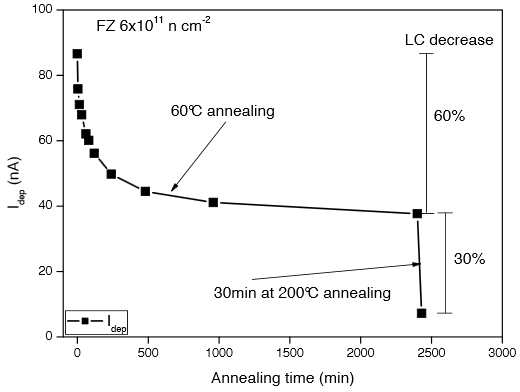}
\caption{Leakage current (LC) decrease in an irradiated float zone (FZ) detector during an annealing at $60^{\circ}$ C and $200^{\circ}$ C. A major part of the initial LC
decreases during the $60^{\circ}$ C annealing for 2400 min and additional 30$\%$ during the 30 minutes at $200^{\circ}$ C \cite{Junkes2011}.}
\label{fig:3}
\end{figure}

The radiation-induced current increase is proportional to the particle fluence when normalised to the damage generated by neutrons with 1 MeV energy according to the Non Ionising Energy Loss (NIEL) model \cite{Huhtinen2002}. 
On the other hand, the leakage current decreases after annealing and higher annealing temperatures result in further reduction, as shown in figure~\ref{fig:3}. DLTS measurements attribute this to the defects E4/E5 and E205a \cite{JunkesPoS2011}, where the E denotes an electron trap having a donor level in the bandgap and the number 205 is the observed temperature dependence of the defect in the DLTS spectrum.

The trapping of the drifting signal charges in the silicon bulk effectively reduces the charge collection efficiency (CCE) of the sensor.

\begin{figure}[tbp] 
\centering 
\includegraphics[width=.45\textwidth]{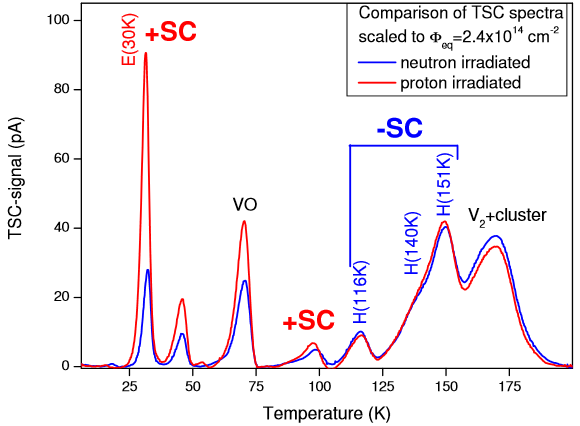}
\caption{TSC spectra measured on the n-type oxygen enriched epitaxial silicon diodes after a proton or neutron irradiation and 30 000 minutes of annealing at $80^{\circ}$ C and $V$ = 100 V. Defects with an impact on
$N_{\textrm{\tiny eff}}$ are marked with +SC for the donors and -SC for the acceptors \cite{Junkes2011}.} 
\label{fig:5}
\end{figure}

The change of the effective doping concentration proceeds in general through the processes of donor removal and acceptor generation. Typically it has been
observed as the device becoming more and more p-type, i.e. a p-type sensor
would stay p-type with an increasing $N_{\textrm{\small eff}}$, and a n-type sensor would additionally undergo an initial
type inversion from n to p. However, measurements of radiation effects on the effective doping concentration in oxygen-enriched n-type epitaxial silicon show a strong dependence on the particle type \cite{Pintilie2009}, demonstrating a significant difference between the results from 23 GeV proton and 1 MeV neutron irradiations, thus violating the simple NIEL scaling hypothesis. 
It is evident in the TSC spectra presented in figure~\ref{fig:5} that the proton irradiation produces a much
higher concentration of the E(30K) defect compared to the neutron irradiation. Differences for the other defects are much less significant. The E(30K) defect has a donor level in the upper part of the band gap, and it generates a positive space charge (SC) in the silicon bulk. $N_{\textrm{\small eff}}$ extracted from the C-V measurements confirms that increased donor concentration in proton irradiated samples prevents the type inversion, while the acceptor-like states dominating in neutron irradiated samples lead to the type inversion before the fluence of $10^{15}$ $\textrm{n}_{\textrm{\small eq}}$cm$^{-2}$ \cite{Parzefall2012}. 

Also electrons in the energy range of 1.5 - 27 MeV have been observed to generate significant concentrations of E(30K) defects with an introduction rate three times larger in the oxygen-enriched silicon material \cite{RD50sr2010,Radu2014}. The origin of the different radiation effects between particles is not yet fully understood, and constitutes an area of ongoing investigation. 

Hence, in the standard n-type (p-type) float zone Si the type inversion occurs (no type inversion occur) irrespective to the type or energy of the radiation, but in the exotic materials like the oxygen-rich Epi-Si the generation of the defects in the band gap depends on the incident particle energy and type. Information of the defects is used for device engineering and as an input to the simulations.
%

% \section{Bibliography styles}
\section{Device Simulations}
\label{Device Simulations}
The information of the defects discussed in the previous section is used as an input to the device simulations. The objective is to develop an approach to model and predict the performance of the irradiated silicon detectors (diode, strip, pixel, 3D) using professional software. Simulations are essential in the e.g. device structure optimization or in predicting the electric fields and trapping in the silicon sensors.

The simulation of radiation damage in the silicon bulk is based on the effective mid gap levels (a deep acceptor and a deep donor level with activation energies $E_{\textrm{a}}$ = $E_{\textrm{c}}$ - (0.525 $\pm$ 0.025) eV and $E_{\textrm{v}}$ + 0.48 eV, respectively). The model was first proposed in 2001 and entitled later as the "PTI model" \cite{Eremin2011}.  
The main idea of the model is that the two peaks in the E($z$) profile of the both proton and neutron irradiated detectors can be explained via the interaction of the carriers from the bulk generated current with the electron traps and simultaneously with the hole traps. 
The physical meaning behind all later models is the same combination of the hole and electron traps.

\begin{figure}[tbp] 
\centering
\includegraphics[width=.47\textwidth]{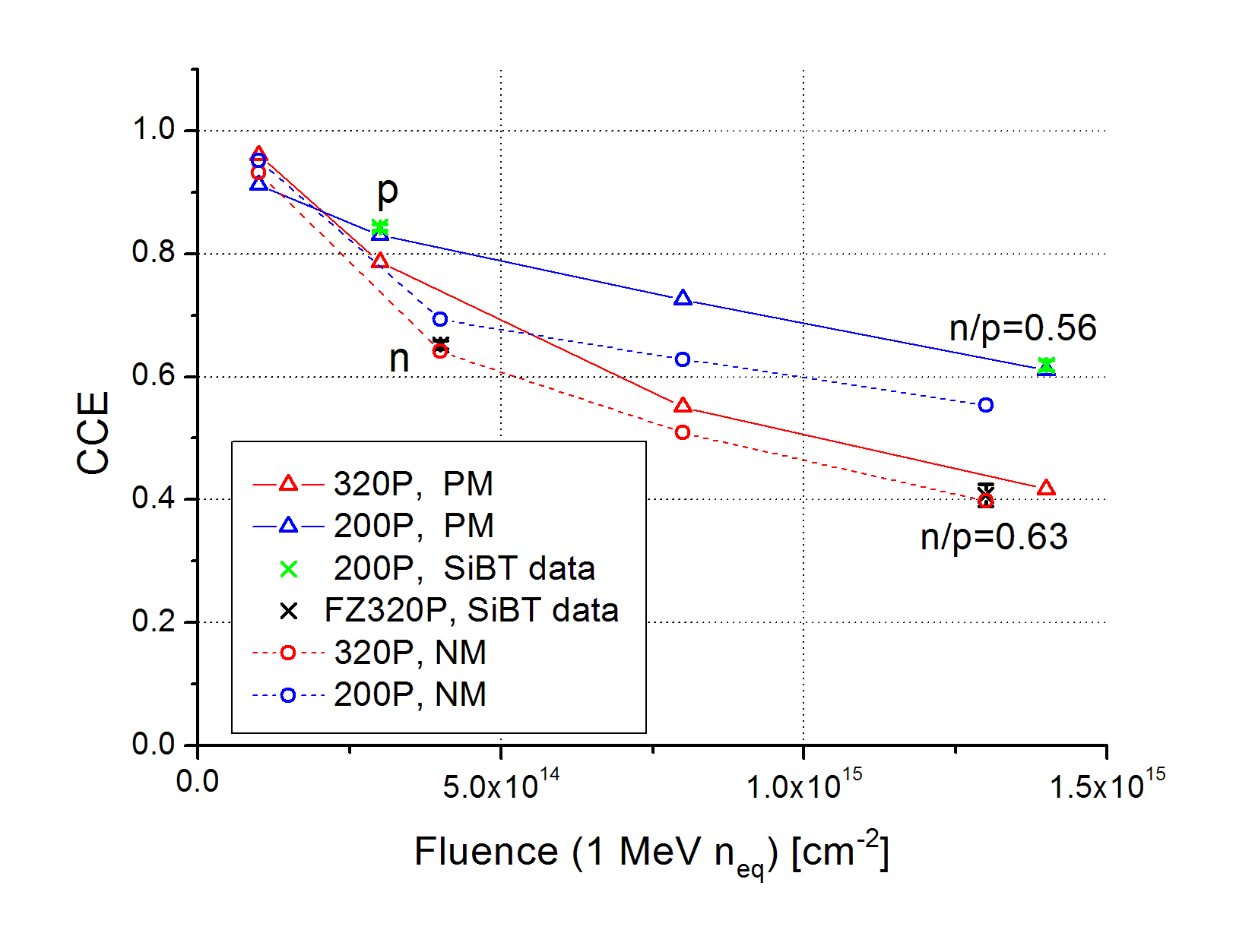}
\caption{Measured and simulated CCE($\Phi_{\textrm{\tiny eq}}$) for the n-on-p strip sensors with a strip pitch of 120 $\mu\textrm{m}$ \cite{Peltola2014}. Types of irradiation are marked in the plot as p = proton, n = neutron and n/p = mixed fluence with the ratios of the particles indicated. PM is the proton model, NM is the neutron model and e.g. 200P is a p-type sensor with a 200 $\mu\textrm{m}$ active region thickness. The experimental data was measured with the SiBT set-up \cite{Harkonen2010}.}
\label{fig:9}
\end{figure}

The first successfully developed quantitative models, namely the proton model and the neutron model \cite{Eber2013}, for the simulation of the detector characteristics like $I_{\textrm{leak}}$, $V_{\textrm{fd}}$ and the charge collection efficiency (CCE), presented in figure~\ref{fig:9}, % , which exploits this concept, 
was built on the base of the PTI model with the same two deep levels as those used for the double peak E($z$) explanation.

\begin{figure}[tbp] 
\centering
\includegraphics[width=.4\textwidth]{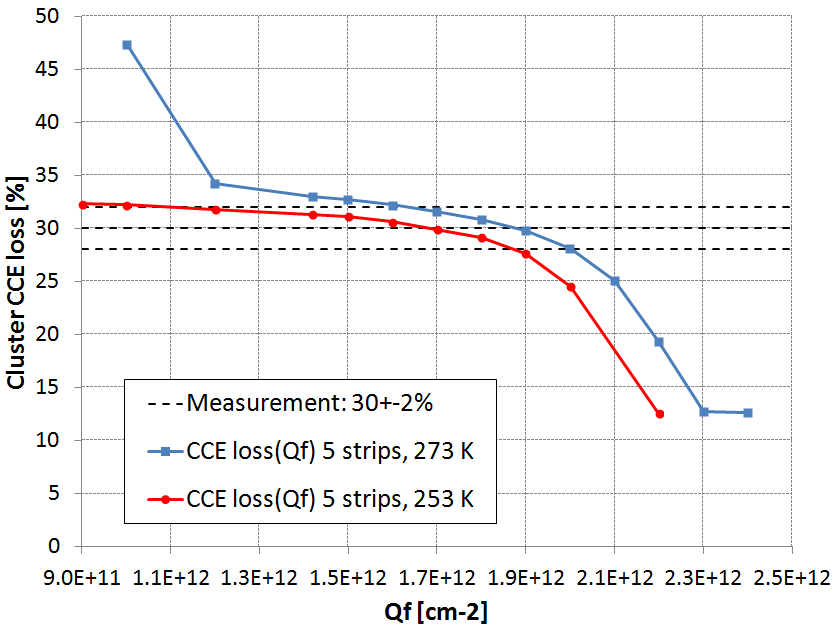}
\caption{Measured and non-uniform 3-level model simulated CCE loss between the strips in a 200 $\mu\textrm{m}$ active thickness n-on-p strip sensor with a 120 $\mu\textrm{m}$ pitch at $\Phi_{\textrm{\tiny eq}}=1.4\times10^{15}$ $\textrm{cm}^{-2}$ and $V$ = -1 kV\cite{Peltola2014}. The agreement with the measurement is given by the interface charge values $Q_{\textrm{\tiny f}}=(1.6\pm0.2)\times10^{12}$ $\textrm{cm}^{-2}$. 
The detectors measured with the SiBT set-up were irradiated by mixed fluences.} 
\label{fig:11}
\end{figure}

The recent implementations of additional traps at the SiO$_2$/Si interface or close to it have expanded the scope of the simulations to include experimentally agreeing surface properties such as the interstrip resistance and capacitance, and the position dependency of CCE, shown in figure~\ref{fig:11}, of the strip sensors irradiated up to $1.5\times10^{15}$ $\textrm{n}_{\textrm{\small eq}}$cm$^{-2}$ \cite{Peltola2014,Peltola2014_2,SimuWG2014,Dalal2014,Eichhorn2014}.
Currently a common database with the cross sections and concentrations is in preparation and more tuning to the data is ongoing.

\section{Detector Characterization}
\label{Detector Characterization}
%%%%%%%%%%%%%%%%%%%%%%%%%%%%%%%%%%%%%%%%%%%%%%%%%%%%%%%%%%%%%%%%%%
\subsection{Detectors on p-type silicon}
\label{Detectors on p-type silicon} 
The planar p-type (or n-in-p) detectors have been brought forward by the RD50 collaboration. They are currently the baseline
for the upgrade of the silicon strip tracking detectors at the ATLAS and CMS experiments. 
The n-in-p sensors have a superior performance resulting from a favourable combination of the weighting and electric fields after irradiation due to the absence of type inversion, and the readout at n-type electrodes, enabling the collection of electrons that have three times higher mobility and longer trapping times than holes. Further essential benefit of the p-type strip sensors is the independence of the CCE of the reverse annealing of the effective space charge in highly irradiated detectors \cite{Casse2002p,Casse2006p,Kramberger2002p}. 

Experiments on the oxygenated and standard 280 $\mu\textrm{m}$ thick p-type microstrip detectors have shown that even after a 23 GeV proton fluence of $7.5\times10^{15}$ cm$^{-2}$
the signal over noise value is $\sim$7.5 which is still a reasonable value for tracking \cite{Casse2004p,Moll2006}.  
%%%%%%%%%%%%%%%%%%%%%%%%%%%%%%%%%%%%%%%%%%%%%%%%%%%%%%%%%%%%%%%%%%
\subsection{Float zone and MCz detectors}
\label{Float zone and MCz detectors}
To improve the radiation tolerance of the p-type detectors, both float zone (FZ) and magnetic Czochralski (MCz) sensors have been investigated. CZ and MCz substrates have, due to their specific growth technique, much higher (about $5\times10^{17}$ cm$^{-3}$) oxygen content than the FZ and even the Diffusion Oxygenated FZ (DOFZ) silicon \cite{Barcz2003}. After being introduced by the RD50 collaboration, it is now one of the most important materials for the detector processing. 

\begin{figure}[tbp] 
\centering
\includegraphics[width=.45\textwidth]{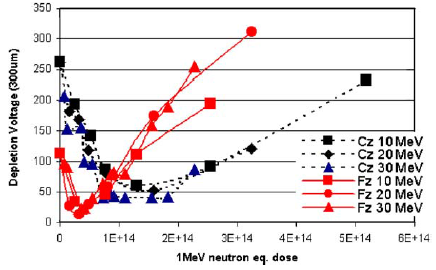}
\caption{Evolution of the full depletion voltage with fluence for the 300 $\mu\textrm{m}$ thick Cz-Si and FZ-Si detectors irradiated with 10, 20 and 30 MeV protons \cite{Harkonen2005}.} 
\label{fig:14}
\end{figure}

Measurements have revealed several advantages in the application of the MCz or Cz substrates. Presented in figure~\ref{fig:14} the type inversion in the n-type Cz-Si has not occurred before $1.6\times10^{14}$ $\textrm{n}_{\textrm{\small eq}}$cm$^{-2}$ fluence. Furthermore, the full depletion voltage in Cz-Si has not exceeded its initial value even after the fluence of $5\times10^{14}$ $\textrm{n}_{\textrm{\small eq}}$cm$^{-2}$ that equals about 5 years of HL-LHC operation for the strip detectors closest to the vertex \cite{Harkonen2003Cz,Harkonen2005}. 
Also the results for the 200 $\mu\textrm{m}$ thick FZ and MCz p-type sensors after mixed irradiation to $1.5\times10^{15}$ $\textrm{n}_{\textrm{\small eq}}$cm$^{-2}$ give a higher collected charge for the MCz sensors  \cite{Steinbruck2013}. The MCz sensors are less affected by the annealing and show a stable annealing behaviour. Also an improved performance in mixed fields due to the compensation of neutron and charged particle damage
in MCz has been observed \cite{Casse2009Cz}.
%%%%%%%%%%%%%%%%%%%%%%%%%%%%%%%%%%%%%%%%%%%%%%%%%%%%%%%%%%%%%%%%%%
\subsection{Thin detectors}
\label{Thin detectors}
\begin{figure}[tbp] 
\centering
\includegraphics[width=.4\textwidth]{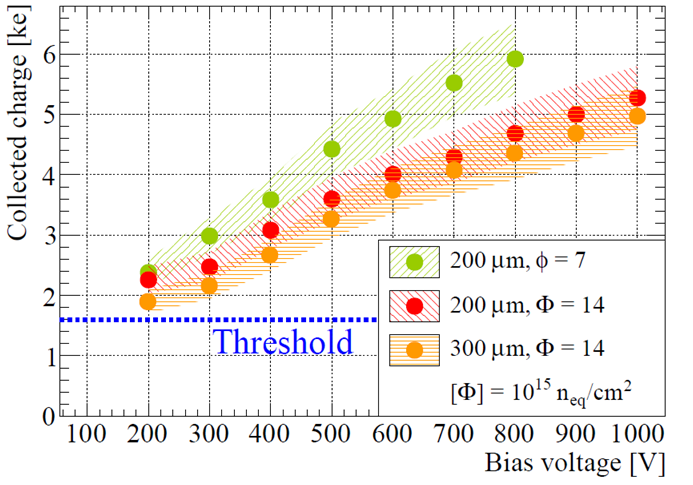}
\caption{The collected charge as a function of the bias voltage for different thicknesses and 800 MeV proton irradiation fluences. At the highest fluence of $14\times10^{15}$ $\textrm{n}_{\textrm{\tiny eq}}\textrm{cm}^{-2}$ the collected charge of the 200 $\mu\textrm{m}$ thick sensors is slightly higher than for the 300 $\mu\textrm{m}$ thick, but still compatible within the estimated uncertainties \cite{Terzo2014}.}
\label{fig:15}
\end{figure}

Benefits of the thin detectors are based on reduced mass and, due to the lower trapping probability, some possible advantage in charge collection and reverse current after high irradiation fluences. While the smaller sensitive volume reduces the signal at low fluences, after high fluences of hadron irradiation and operated above $V_{\textrm{fd}}$ the thin detectors could yield a higher signal than thick detectors that cannot reach full depletion because of the high value of $N_{\textrm{eff}}$ \cite{Casse2009Cz}. Presented in figure~\ref{fig:15} are the collected charges in the 200 and 300 $\mu\textrm{m}$ sensors after extremely high proton fluences.

\section{New Structures}
\label{New Structures}
The investigations of new structures cover, besides the above discussed thin detectors, a variety of new devices like the different types of 3D detectors, sensors with intrinsic gain and slim or active edge
sensors.
%%%%%%%%%%%%%%%%%%%%%%%%%%%%%%%%%%%%%%%%%%%%%%%%%%%%%%%%%%%%%%%%%%%
\subsection{3D Detectors}\label{sec:5a}
The 3D detectors are likely to be the most promising choice for the extremely high fluence environments. By having doped columns etched into the silicon bulk vertically to the device surface, the geometry of the 3D sensors decouples the depletion voltage and the detector thickness. This means that also the charge drift length is decoupled from the ionization path, i.e. the drift length is now the inter-column spacing, while the signal is still proportional to the detector thickness resulting in a higher radiation tolerance \cite{Parker1997}. 

The downside of the 3D sensors is a more complex processing, a higher inter-electrode capacitance (with higher electronics noise), non-sensitive regions in the columns and the hit position dependent signal size. The disadvantage of the low-field regions is overcome by the 3D trench electrode detectors that have a concentric trench electrode
surrounding the central hexagonal signal collecting column electrode \cite{Li2011,Montalbano2014}. The hexagonal electrode design results in an uniform electric field.

In addition to the single column type \cite{Piemonte2005}, also a double column-double sided technology has been developed by the RD50 institutes \cite{Pellegrini2008}. Due to being available in large production and being able to give an acceptable signal after irradiation to fluences of $1\times10^{16}$ $\textrm{n}_{\textrm{\small eq}}\textrm{cm}^{-2}$, the 3D sensors now populate 25$\%$ of the ATLAS experiment insertable b-layer (IBL) \cite{ATLASr2010}.

\begin{figure}[tbp] 
\centering
\includegraphics[width=.4\textwidth]{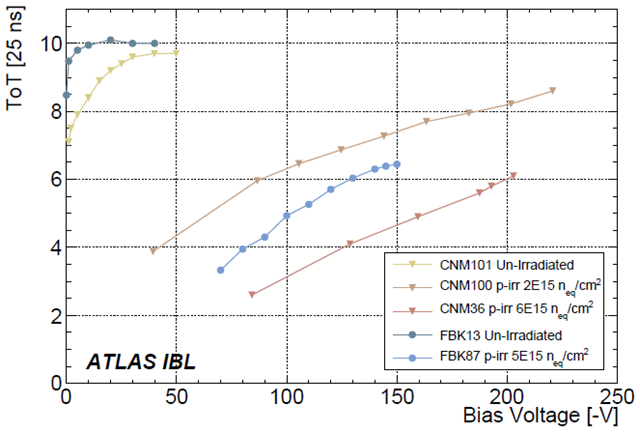}
% \begin{minipage}{\linewidth}
\caption{The most probable value (MPV) of the time over threshold (ToT) spectrum measured for selected FBK and CNM 3D modules using a $^{90}$Sr source. The data are shown in units of the 25 ns bunch crossing clock as a function of $V_{\textrm{b}}$. Irradiated modules at several fluences are compared with non-irradiated samples \cite{ATLASibl2012}.} 
% \end{minipage}
\label{fig:16}
\end{figure}

Presented in figure~\ref{fig:16} is a study of charge collection in the irradiated FBK\footnote{Fondazione Bruno Kessler, www.fbk.eu} and CNM\footnote{Centro Nacional de Microelectr\'{o}nica, www.cnm.es} produced 3D modules. Since the expected maximum fluence at the IBL is $5\times10^{15}$ $\textrm{n}_{\textrm{\small eq}}\textrm{cm}^{-2}$, the results indicate that operation at $V_{\textrm{\small b}}\geq$ 160 V is needed for an acceptable charge collection efficiency after irradiation \cite{ATLASibl2012}.
%
% \footnotetext{Fondazione Bruno Kessler, www.fbk.eu}
% \footnotetext{Centro Nacional de Microelectr\'{o}nica, www.cnm.es}
%%%%%%%%%%%%%%%%%%%%%%%%%%%%%%%%%%%%%%%%%%%%%%%%%%%%%%%%%%%%%%%%%%%
\subsection{Active and slim edge detectors}\label{sec:5b} 
Reduced edge width is a method used to maximize the sensitive region of the silicon sensors in the tracking detectors. The first edgeless detectors using a break-through junction and operated at low $T$ (130 K) were studied in \cite{Li2002}. Later a new construction was proposed based on the optimization of the electric field distribution on the edge and its control by the amorphisation of the edge surface \cite{Ruggiero2009} making useless the voltage termination structure between the edge and the detector active area. This structure can be totally replaced by a single ring which controls the current distribution between the edge and the sensitive volume. This entirely solves the problem with the edgeless detector premature breakdown in detectors with a very narrow 45 $\mu\textrm{m}$ non-sensitive contour at the edge. On the basis of this principle 240 microstrip detectors out of 400 processed were installed and successfully operated at the TOTEM experiment during the first two-year long run of the LHC.

\begin{figure}[tbp] 
\centering
\includegraphics[width=.4\textwidth]{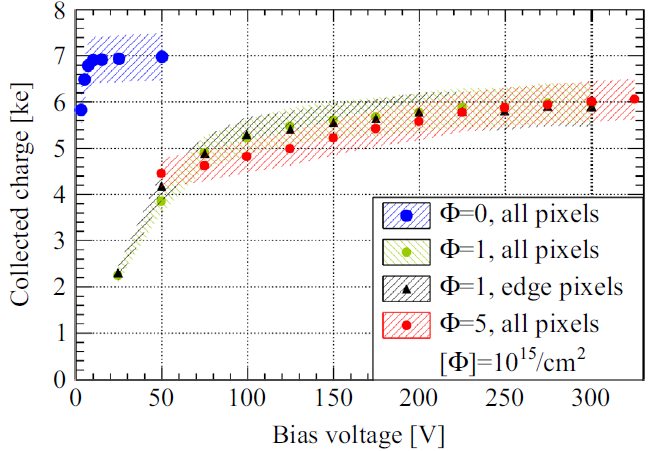}
\caption{Collected charge of a VTT FE-I3 module, with an edge width of 125 $\mu\textrm{m}$, irradiated first with 25 MeV protons at $\Phi=10^{15}$ n$_{\textrm{\tiny eq}}\textrm{cm}^{-2}$ and then with reactor neutrons at $\Phi=4\times10^{15}$ n$_{\textrm{\tiny eq}}\textrm{cm}^{-2}$. The pixel threshold was tuned to 1500 electrons for all these measurements \cite{Macchiolo2014}.} 
\label{fig:19}
\end{figure}

Further approaches have since been developed to minimize the non-active volume, resulting in slim and active
edge sensors. The scribe-cleave-passivate technology aims to reach 100 $\mu\textrm{m}$ or less slim edges, while the typical distances to the edge
are in the range 500-1000 $\mu\textrm{m}$ \cite{Fadeyev2013}. Alternatively, pixel sensors with
active edges are manufactured by wet etching trenches at the sensor border that are then activated by the means of four-quadrant ion implantations \cite{Macchiolo2014}. Presented in figure~\ref{fig:19} are the charge collection measurements of the thin p-type pixel sensors with 125 $\mu\textrm{m}$ width active edges (processed using the last mentioned method at VTT\footnote{VTT Technical Research Centre of Finland Ltd., http://www.vtt.fi}) before and after irradiation. Similar median charge for the strips close to the edge and other strips are collected. 
% 
%%%%%%%%%%%%%%%%%%%%%%%%%%%%%%%%%%%%%%%%%%%%%%%%%%%%%%%%%%%%%%%%%%%
\subsection{Sensors with intrinsic gain}\label{sec:5c}
The effect of the charge multiplication has been observed in several device types, namely the strip sensors, 3D sensors and diodes \cite{Lange2010,Casse2010cm,Koehler2011}. The
enhancement arises from the carrier avalanche multiplication in the high electric field of the junction, resulting from a high negative space charge concentration in the bulk after irradiation \cite{Verbitskaya2013}. 
 
Extensive modelling of the detectors with intrinsic gain has already been carried out \cite{Verbitskaya2012,Verbitskaya2013} and a dedicated program has been setup within the RD50
collaboration to understand the underlying mechanisms of the multiplication, to simulate them and to optimize the CCE performances.
% to understand and optimize the mechanism and to evaluate its exploitation. 

To this end special charge multiplication sensors have been fabricated. 
The newly developed sensors with intrinsic gain, i.e. Low Gain Avalanche Detectors (LGAD), are based on avalache photodiode technology and feature an implemented multiplication
layer, a deep p$^+$ implant below the cathodes. An edge termination done by a low doping n-well is needed to secure gain uniformity \cite{Pellegrini2014}. 

\begin{figure}[tbp] 
\centering
\includegraphics[width=.47\textwidth]{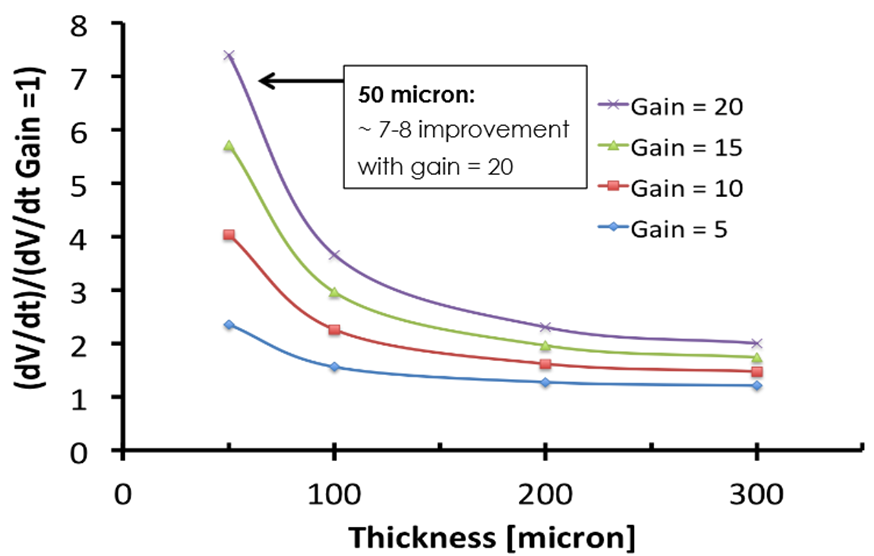}
\caption{Simulation of the signal time differential as a function of the device thickness for sensors with varying intrinsic gain. Significant improvements in time resolution are observed only for the thin detectors  \cite{Cartiglia2014}.} 
\label{fig:22}
\end{figure}

Gain values of up to $\sim$20 for $^{90}$Sr electrons have been measured \cite{Kramberger2014rd50} in the non-irradiated LGADs while the leakage current and noise are independent of the gain \cite{Kramberger2013rd50}. With this technology thinner detectors can be produced to give the signal of thick ones,
enabling the investigation of ultra-fast detectors \cite{Sadro2013,Sadro2014} for HL-LHC. The dependence of the time resolution on gain and device thickness is presented in figure~\ref{fig:22}.
% Also transient current technique measurements show the occurence of the multiplication effect. 

After irradiation a significant reduction in gain is occuring. Neutron irradiation to a fluence of $1\times10^{16}$ $\textrm{cm}^{-2}$ leads to a signal that is compatible with a standard diode and for 800 MeV protons of fluence $\sim5\times10^{15}$ $\textrm{n}_{\textrm{\small eq}}\textrm{cm}^{-2}$
almost no gain can be observed. Instead of trapping the effective acceptor removal in the p$^+$ layer is responsible for the gain degradation by reducing the electric field  \cite{Kramberger2014rd50}. Presently more wafers are in production for further tests and an irradiation programme has been planned.

\section{Summary}
\label{Summary}
An overview of the RD50 collaboration research activities for developing sensors with sufficient radiation hardness for the
HL-LHC has been given including several examples of the work performed by the RD50 collaboration.

Defect characterization results indicate that in the standard n-type (p-type) float zone Si the type inversion occurs (no type inversion occur)
irrespective to the type or energy of radiation, but in more exotic materials like the oxygen-rich Epi-Si the generation of the defects in the band gap depends on the
incident particle energy and type.

TCAD simulations carried out by a dedicated work group are now able to account for both bulk and surface damage resulting in further convergence with the measurements and increasing predictive power.

The sensors with n-electrode readout (mainly sensors with p-bulk) offer the advantage of collecting electrons instead of holes resulting in an improvement of radiation tolerance. For the outer layer with maximum fluences of about $2\times10^{15}$ $\textrm{n}_{\textrm{\small eq}}\textrm{cm}^{-2}$ planar n-in-p sensors are the baseline for the ATLAS and CMS upgrade strip tracker. MCz was shown to be more radiation tolerant than the standard FZ silicon with respect to the change of the effective doping concentration. Furthermore, it has been observed that in the n-MCz material neutron and proton irradiations introduce effects that act as doping of opposite polarity, leading to a possible partial cancellation effect of the degradation of $N_{\textrm{eff}}$. The thin detectors overcome the requirement of high bias voltages and are advantageous at fluences above $10^{15}$ $\textrm{n}_{\textrm{\small eq}}\textrm{cm}^{-2}$. 

The 3D detectors are a promising option and show good performance requiring lower bias voltages. The designs include now the single column type, the double column-double sided type and the trench electrode detectors. Due to being able to give an acceptable signal after irradiation to $1\times10^{16}$ $\textrm{n}_{\textrm{\small eq}}\textrm{cm}^{-2}$, the 3D sensors now populate 25$\%$ of the ATLAS IBL. The edgeless, active or slim edge devices have been developed to maximize the sensitive region of the silicon sensors in tracking detectors. 240 edgeless microstrip detectors were already installed and successfully operated at the TOTEM experiment during the first two-year long run of LHC.
The effect of the charge multiplication is investigated systematically to allow its exploitation and dedicated sensors have been produced. The sensors with intrinsic gain offer the possibility for the thin detectors to produce the signal of thick ones, enabling the investigation of ultra-fast planar detectors for HL-LHC. This requires first a solution to the significant reduction of gain after irradiations at expected fluences.
 
For further results aiming for radiation hard semiconductor devices for the future colliders, the website of the RD50 collaboration is referred.

\section*{Acknowledgement}
The author would like to thank the colleagues of the RD50 Collaboration for the material and support.

% \section*{References}

\bibliography{mybibfile}

\end{document}